\documentclass[aps,pra,preprint,groupedaddress,showpacs]{revtex4-1}

\usepackage{subfigure} 
\usepackage{graphicx}  
\usepackage{dcolumn}   
\usepackage{amsmath}
\begin{document}
\preprint{Version 3.0}

\title{Field-theory calculation of the electric dipole moment of the neutron 
and paramagnetic atoms}

\author{S. A. Blundell}
\email{steven.blundell@cea.fr}
\affiliation{SPSMS, UMR-E CEA/UJF-Grenoble 1, INAC, Grenoble, F-38054, France.}

\author{J. Griffith}
\email[]{jgriffi8@nd.edu}
\affiliation{Department of Physics,
University of Notre Dame, Notre Dame, IN 46556}

\author{J. Sapirstein}
\email[]{jsapirst@nd.edu}
\affiliation{Department of Physics,
University of Notre Dame, Notre Dame, IN 46556}

\begin{abstract}
Electric dipole moments (edms) of bound states that arise from the 
constituents having edms are studied with field-theoretic techniques. The 
systems treated are the neutron and a set of paramagnetic atoms. In the 
latter case it is well known that the atomic edm differs greatly from the 
electron edm  when the internal  electric fields of the atom are taken 
into account. In the nonrelativistic limit these fields lead to a complete 
suppression, but for heavy atoms large enhancement factors are  present. 
A general bound-state field theory approach applicable to both the neutron and 
paramagnetic atoms is set up. It is applied first to the neutron, treating 
the quarks as moving freely in a confining spherical well. It is shown that 
the effect of internal electric fields is small in this case. The atomic 
problem is then revisited using field-theory techniques in place of the usual 
Hamiltonian methods, and the atomic enhancement factor is shown to be 
consistent with previous calculations. Possible application of bound-state 
techniques to other sources of the neutron edm is discussed.
\end{abstract}
\pacs{13.40.Em,12.39.Ba,11.30.Er}

\maketitle

\section{introduction}

One of the strongest constraints on new physics beyond the standard model is 
provided by measurements attempting to detect a nonvanishing electric dipole 
moment (edm) of the neutron, heavy atoms, or molecules \cite{Roberts}. This 
is because the extremely small value of the edm of quarks and the electron in 
the standard model means that any detection with presently available 
sensitivities arises from such new physics.  According to Schiff's theorem \cite{Schiff}, the edm of atoms or molecules arising from an electron edm vanishes in the nonrelativistic limit.  However, the power of experiments on heavy atoms or molecules to put limits on a possible electron edm $d_e$ is in fact increased by orders of magnitude because of a large violation of Schiff's theorem discovered by Sandars \cite{Sandars}, which stems from relativistic effects. Typical values of the enhancement factor $R$, defined 
by
\begin{equation}
  d_{\rm{atom}} = R d_e,
  \label{Rdef}
\end{equation}
are $R=120$ for cesium \cite{cs} and $R=-685$ for thallium \cite{tl}. These 
calculations of $R$ are carried out in a Hamiltonian formalism, with the 
relativistic many-body problem treated using many-body perturbation theory 
(MBPT), generally summing infinite classes of MBPT diagrams for higher 
accuracy.

One of the purposes of the present paper is to recalculate the enhancement 
factor for paramagnetic atoms, specifically thallium and the alkali metals 
lithium through francium, using a different formalism, one which treats 
corrections in terms of one photon exchange. Our aim here is not to achieve 
high accuracy in the many-body part of the problem, but rather to formulate 
the calculation in a field-theoretic framework \cite{Lindroth}. This 
framework allows us to address the main purpose of the paper, calculating the 
neutron edm induced when the up and down quarks have edms $d_u$ and $d_d$ 
respectively. We are specifically concerned with the question of how the 
internal electric fields arising from the charged quarks affect the edm of 
the bound state, since their role is so important in the atomic case.  In 
order to apply the techniques of atomic physics to the neutron, we treat the 
up and down quarks as moving freely in a confining well. 
This is a simplified version of the MIT bag model \cite{MIT},  without 
complicating issues such as bag pressure and the nonlinear boundary condition; 
we also neglect gluon exchange. We refer to this approach as the static well 
model in the following. The model is known to give the neutron (and proton) magnetic moment to within 20 percent of the experimental value when the quarks are treated as nearly massless particles with magnetic moments coming from their confinement in the well \cite{Lee}, so using it to evaluate the neutron edm that arises from 
nonvanishing values of $d_u$ and $d_d$ should give results of similar 
accuracy.

The paper is organized as follows. In the next section we introduce the bound-state field theory tools that will be used for both the neutron and 
paramagnetic atom calculations, both of which are variations of the Furry representation 
\cite{Furry}.  In the following section we calculate first the lowest-order 
edm of the neutron, finding a 17 percent reduction from the standard 
nonrelativistic result, and then evaluate the effect of one photon exchange between quarks, which is equivalent to calculating the effect of the internal electric fields in the neutron. This effect will be shown to be quite small, so that despite the highly relativistic nature of the quarks within the model, we do not find large edm enhancement factors for the neutron.  We then turn to the atomic problem, beginning with a brief description of the violation of Schiff's theorem \cite{Schiff} in atomic physics for paramagnetic atoms, after which we present a field-theoretic treatment of the problem analogous to that used for the neutron.  We show that the approach leads to 
a suppression for light alkalis along with the well-known large enhancement factors for heavy atoms.  In the concluding section we discuss how different sources of CP nonconservation could be treated with 
the methods described here, and also ways in which the atomic calculations could be improved.

Because we will be dealing with elementary particles with different electric 
charges, in the following we will present all formulas with factors of $q_a$, 
with $q_e=-e$, $q_u = 2/3e$, and $q_d = -1/3e$, $e$ taken to be positive. 

\section{Bound-State Field Theory}

For both problems treated here we assume spherical symmetry, and we expand 
field operators in terms of solutions to the Dirac equation 
\begin{equation}
  \psi_n(\vec r) = \frac{1}{r} \left( \begin{array}{c} ig_n(r) 
  \chi_{\kappa \mu}(\hat{r}) \\
  f_n(r) \chi_{-\kappa \mu}(\hat{r}) \end{array} \right),
  \label{Dirac}
\end{equation}
with energy eigenvalues $\epsilon_n$, and where
$\chi_{\kappa \mu}$ is a spherical spinor. To treat the bound state problem we 
use variants of the Furry representation, originally introduced 
to evaluate QED effects in hydrogen. This representation can be thought of as 
intermediate between the standard interaction and the Schr\"{o}dinger 
representations. 

To transform from the Schr\"{o}dinger to the interaction representation the unitary 
transformation 
\begin{equation}
  |\phi_I(t) \rangle = e^{i H_0 t} |\phi_S(t) \rangle
\end{equation}
is carried out, with
\begin{equation}
  H_0 = \int d^3 x \psi^{\dagger}(x)
  [\vec \alpha \cdot \vec p + \beta m   ] \psi(x).
\end{equation}
In the Furry representation, the unitary transformation is
\begin{equation}
  |\phi_F(t) \rangle = e^{ i \tilde{H}_0 t} | \phi_S(t) \rangle.
\end{equation}
where, defining $r=|\vec x|$,
\begin{equation}
\tilde{H}_0 = \int d^3 x \psi^{\dagger}(x)
  [\vec \alpha \cdot \vec p + \beta m(r)  - {Z \alpha \over r} ] \psi(x).
\end{equation}
We have replaced the standard electron mass term $\beta m$ with a 
position-dependent mass for later generalization to the static well model, 
though it is of course constant in atomic physics. This representation is used 
to this day to treat radiative corrections in hydrogen and hydrogenic ions, 
where the primary role of the nucleus is to provide a classical Coulomb field
centered at the origin, which introduces the extra term into $\tilde{H}_0$. 
(Finite nuclear size effects, which are particularly important for highly 
charged ions, are accounted for by taking $Z \rightarrow Z(r)$, with $Z(r)$ 
modeled by a Fermi distribution).

After the transformation to the Furry representation the interaction Hamiltonian (for the atomic problem) has 
the usual form
\begin{equation}
  H_I = q_e \int d^3 x \bar{\psi}(x) \gamma_{\mu} \psi(x) A^{\mu}(x) 
\end{equation}
with $A^{\mu}$ the quantized radiation field, but with the electron field operators 
now expanded in terms of solutions to the Dirac equation in an external 
Coulomb field of the form given in Eq. \eqref{Dirac}.

To evaluate energy corrections, a generalization of the Gell-Mann--Low approach 
introduced
by Sucher \cite{Sucher} can be used,
\begin{equation}
  \Delta E = \lim_{\epsilon \rightarrow 0, \lambda \rightarrow 1} 
  {i \epsilon \over 2} 
  {\partial \over \partial \lambda} \ln \langle \phi |\,
  T( e^{-i \lambda \int dx_0 e^{-\epsilon |x_0|} H_I(x_0)})| \phi \rangle,
\end{equation}
which has an $S$-matrix modified by
the factor $e^{-\epsilon |x_0|}$, included in order to adiabatically turn 
off the interaction at large positive and negative times. This allows standard Feynman 
diagram techniques to be applied, with the adiabatic factor usually leading trivially to a 
factor $1/\epsilon$ that cancels the $\epsilon$ in the numerator of the above 
formula.

In this work we are interested in treating many-electron atoms and the neutron, but only 
slight modifications of the approach used for hydrogen described above are needed. We 
start with the neutron. We model it as three light 
quarks confined in a spherical well of radius $R=1.2$ fm, choosing the scalar mass 
of the quarks to be 
finite inside the well and infinite outside, which leads to the boundary 
conditions of the MIT bag model \cite{MIT}. Just as with the description of 
hydrogen, we note that there is a fixed origin, in this case the center of 
the well instead of the position of the nucleus. In both cases this is 
equivalent to neglecting recoil effects. Because the masses of the up and down 
quarks we will be considering are very light, we set them equal to zero. The 
calculations presented below were tested with nonzero quark masses and found 
to be insensitive to this approximation. A simplification of choosing the up 
and down quark masses to be equal is that the spatial wave functions are then 
identical.

For the neutron, then, we take
\begin{equation}
  \tilde{H}_0 = \sum_i \int d^3 x [{\psi_{ui}}^{\dagger}(x)
  [\vec \alpha \cdot \vec p + \gamma_0 m(r)] {\psi_{ui}}(x) 
   + {\psi_{di}}^{\dagger}(x)
  [\vec \alpha \cdot \vec p + \gamma_0 m(r)] {\psi_{di}}(x)]
\end{equation}
where $i$ is a color index. In this case the quark fields are 
expanded in terms of solutions of the free Dirac equation for massless 
particles with MIT bag model boundary conditions. The neutron wave function 
consists of a product of three $1s$ wave functions \cite{Lee}, with 
\begin{equation}
  |\,n_{\uparrow} \rangle  = {\epsilon_{ijk} \over 2\sqrt{18}}[-2 b^{\dagger}_{ib} 
  b^{\dagger}_{jc} b^{\dagger}_{kc}
  + b^{\dagger}_{ia} b^{\dagger}_{jd} b^{\dagger}_{kc} 
  +  b^{\dagger}_{ia} b^{\dagger}_{jc} b^{\dagger}_{kd}]|0 \rangle,
  \label{neutronwf}
\end{equation}
where we have introduced a notation in which $a$ and $b$ denote spin up and 
down states of an up quark, and $c$ and $d$ spin up and down states of a down 
quark. The color indices $i,j,$ and $k$ are again understood to be summed 
over. The ground state energy of a massless quark when $R=1.2$~fm is 335.9 MeV, which will
be denoted $\epsilon_g$ in the following: we also denote the associated wave function as
$\psi_g(\vec x)$.
This defines the starting point for our treatment of the neutron edm, and 
calculations involving the appropriate interaction Hamiltonians can now be 
carried out in a systematic manner.

We now turn to the atomic case, where we will be interested in treating atoms 
with many electrons, in which case the original Furry representation is  
inadequate. To approximate the effects of screening, we introduce a local, 
central potential chosen to be close to the Hartree-Fock (HF) potential. Most 
atomic calculations based on Hamiltonian methods in fact use the latter 
potential, but its nonlocality makes it unsuitable for field theory 
calculations. The new free Hamiltonian is
\begin{equation}
  \tilde{H}_0 = \int d^3 x  {\psi}^{\dagger}(x)
  [\vec \alpha \cdot \vec p + \beta m - {Z \alpha \over r} + U(r)] 
  \psi(x).
\end{equation}
Here we use a potential, which we call the core-Hartree (CH) potential, 
defined by 
\begin{equation}
  U_{\rm CH}(r)= \alpha \int_0^\infty \! {dx\over r_m(x)} \rho(x),
\end{equation}
where $r_m(x)=$max$(r,x)$ and 
\begin{equation}
  \rho(r) = \sum_{a}(g^2_a(r) + f^2_a(r))
\end{equation}
is the sum of the charge densities of all electrons in the filled core. This
potential gives results close to the HF potential. To illustrate 
the accuracy of the potential we note that for cesium it predicts the removal 
energy of the valence $6s$ electron to be $3.267$ eV, to be compared with 
$3.466$ eV for the HF potential result. (The experimental value is 
$3.894$ eV.) Because we have incorporated this potential into $\tilde{H}_0$, 
the introduction of a new interaction, which acts like a counterterm,
\begin{equation}
  H_{\rm CT} =  -\int d^3 x \bar{\psi}(x) \gamma_0 \psi(x) U_{\rm CH}(r),
\end{equation}
is required. At the level of perturbation theory we use in this paper 
it plays no role. 

By working with this modified Furry representation, it is possible to put 
atomic many-body calculations in a field-theoretic framework, with Feynman 
diagrams involving Coulomb photon exchange and $H_{\rm CT}$ having a direct 
correspondence with the formulas of MBPT \cite{Rev}.  Physics associated with 
transverse photon exchange and negative-energy states enters in a well-defined 
manner. The issue of negative-energy states, which require care to properly 
include in Hamiltonian formalisms, is of course correctly treated in a field-theoretical approach. This issue was emphasized in Ref. \cite{Lindroth}.

To treat the many electrons present (up to 87 for francium), we exploit the
fact that all atoms considered here have one electron outside a filled core,
denoted $| O_C \rangle$. Standard methods from many-body perturbation theory
can then be used to treat the paramagnetic ground states as
\begin{equation}
  | \phi \rangle = b^{\dagger}_{v} | 0_C \rangle,
\end{equation}
where $v$ denotes a valence electron outside a closed-shell 
core. For example, cesium would have $v$ being the $6s$ state, and
$|0_C \rangle$ a xenon-like core, with the 54 electrons filling the $1s-5s$, 
$2p-5p$, and $3d-4d$ shells. The completeness of the shells allows angular 
momentum identities to simplify the calculations, but summations over all the 
core states are still needed. 

\section{Quark edm contribution to $d_n$}

We confine our attention here to the contribution of nonvanishing edms of the 
up and down quark to the neutron edm. A thorough discussion of other 
contributions to the overall edm of the neutron is given in Chapter 13 of Ref. \cite{Roberts}.
We will present results for the lowest order contribution, $d_n^{(1)}$, and for
the correction due to internal electric fields, $d_n^{(2)}$.
 
A general spin-1/2 particle with edm $d$ is described by the interaction 
Hamiltonian
\begin{equation}
  H_{\rm edm} = i {d \over 2} \int d^3 x \bar{\psi}(x)\sigma_{\mu \nu} 
  \gamma_5 \psi(x) F^{\mu \nu}(x),
  \label{edm0}
\end{equation}
where $F^{\mu \nu}(x)$ at this point includes both classical and quantized 
fields. We begin by taking the electromagnetic field to be an external 
constant electric field pointing in the positive $z$ direction. A neutron 
with edm $d_n$ in such a field has a linear Stark effect, with the energy 
shift in the rest frame of the neutron when its spin is up being
\begin{equation}
  \Delta E = -d_n E_{\rm{ext}}.
\end{equation}
To calculate the effect  
the edms of up and down quarks bound in a neutron have on $d_n$, we need to evaluate the 
energy shift in the same external electric field using the static well model 
described above, and we will then identify the coefficient of $-E_{\rm{ext}}$ 
as the neutron edm. Because the neutron has both up and down quarks, we 
generalize Eq. \ref{edm0} to
\begin{equation}
  H_{I3} = -id_u E_{ext} \sum_i \int d^3 x \bar{\psi}_{ui}(x) 
  \sigma_{03} \gamma_5 \psi_{ui}(x) 
          -id_d E_{ext} \sum_i \int d^3 x \bar{\psi}_{di}(x) 
  \sigma_{03} \gamma_5 \psi_{di}(x)
\end{equation}
where we have replaced $F^{\mu \nu}$ with its external-field value. It is then
straightforward  to apply the Gell-Mann--Low formalism to calculate the energy 
shift and identify the lowest-order contribution as 
\begin{equation}\label{dn1}
  d_n^{(1)} = 0.8265 [\frac43 d_d - \frac13 d_u].
\end{equation}
The factor 0.8265 is the integral
\begin{equation}
  I = \int_0^R \! dr \, [g_g^2(r) + \frac13 f_g^2(r)]
\end{equation}
for the spherical well model with massless quarks. The same kind of integral, which apart from the factor of 1/3 is the normalization integral, 
is present also in the atomic edm problem for valence $s_{1/2}$ states, but it is extremely close to unity in that case, even for heavy atoms. Here, however, because the quarks are totally relativistic, the integral changes by more than 17 percent from unity. We note that in nonrelativistic quark models the factor is taken to be exactly one. This 17 percent change from the nonrelativistic limit will turn out to dominate by far the effect of the internal electric fields in the neutron. 

We now turn to the evaluation of these effects, which are associated with 
one photon exchange. The motivation, as mentioned above, for evaluating the 
effect of such terms comes from the case of atomic physics. As will be shown
in the next section, in heavy atoms  these terms 
produce large enhancement factors. 
Because this latter enhancement is a relativistic effect, it is difficult to 
form an intuitive understanding of it. We note that while several explanations 
have been given, a recent paper \cite{Commins} raises questions about their 
validity, and gives a quite different derivation involving 
Lorentz-Fitzgerald contraction. It is our view that in such situations there 
is no substitute for an explicit calculation of the effect.

\begin{figure}[b]
  \begin{center}
    \subfigure[]{\label{1a}\includegraphics[width=7cm]{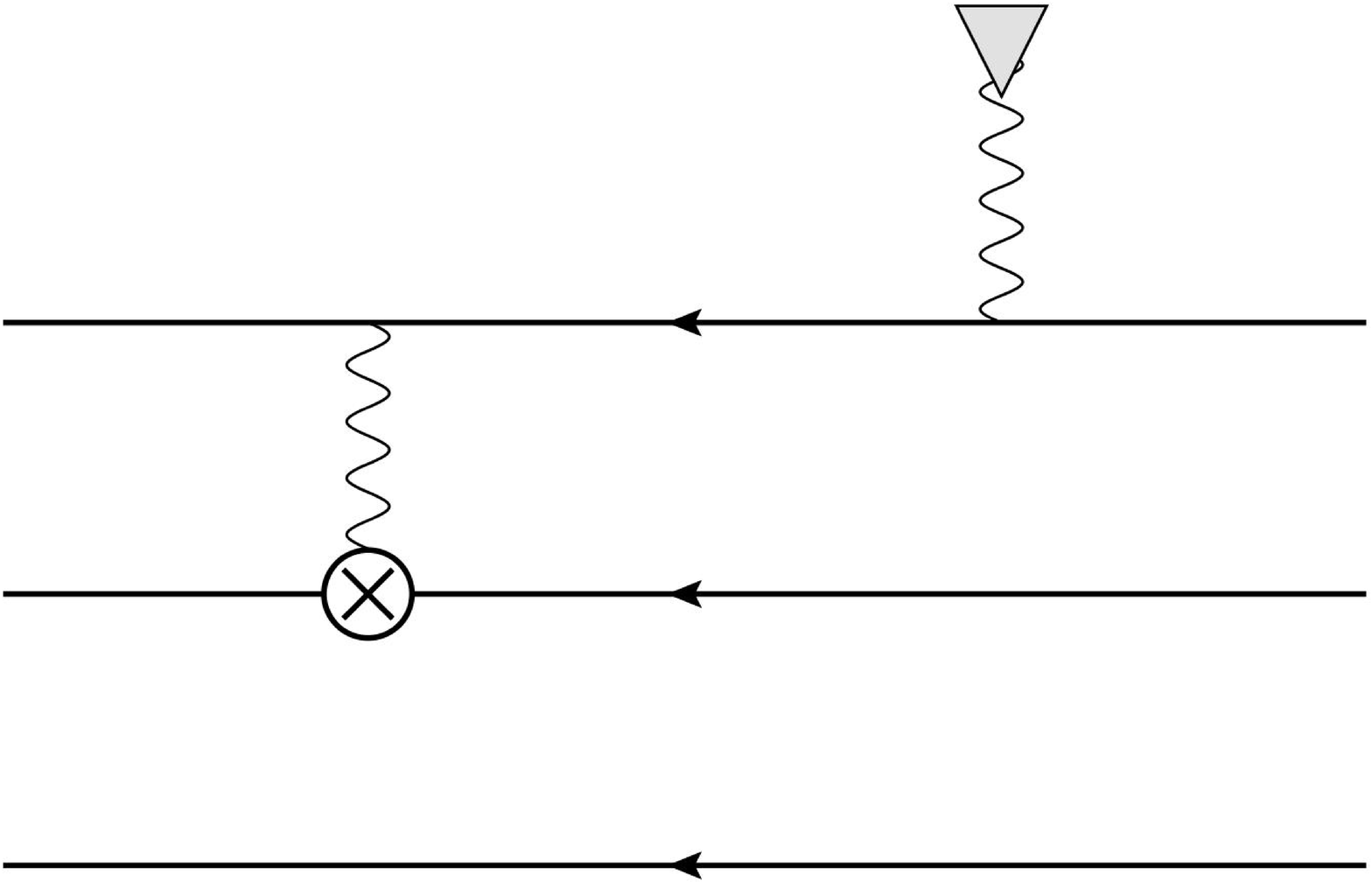}}
    \qquad
    \subfigure[]{\label{1b}\includegraphics[width=7cm]{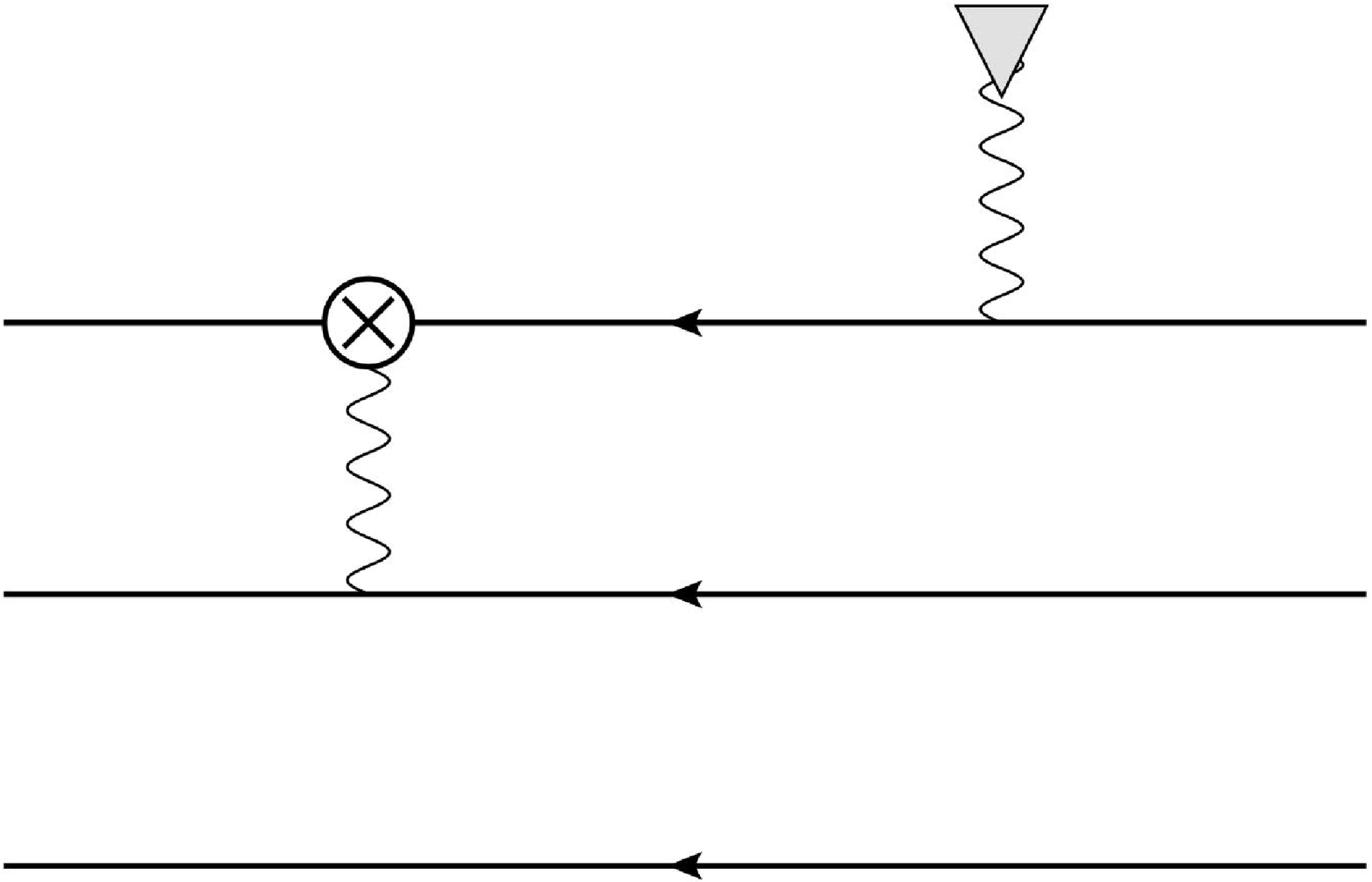}}
  \end{center}
  \caption{The two classes of one-photon-exchange diagrams contributing 
    to the edm of the neutron. 
    The triangle vertex represents the external electric field, while the 
    crossed vertex represents the edm interaction.}
    \label{1ab}
\end{figure}

The electromagnetic field is now split into two parts, the classical field 
used in the lowest-order calculation described above, and a quantized field. 
This leads to three additional Hamiltonians, 
\begin{eqnarray}
  H_{I1} & = & q_u \sum_i  \int \! d^3 x \, \bar{\psi}_{ui}(x) \gamma_{\mu} 
  \psi_{ui}(x) A^{\mu}(x)  
        +  q_d \sum_i  \int \!  d^3 x \, \bar{\psi}_{di}(x) \gamma_{\mu} 
  \psi_{di}(x) A^{\mu}(x) \nonumber \\
  H_{I2} & = & -q_u E_{ext} \sum_i \int \!  d^3 x \, x_3 \bar{\psi}_{ui}(x) 
  \gamma_0 \psi_{ui}(x) 
        -  q_d E_{ext} \sum_i \int \!  d^3 x \, x_3 \bar{\psi}_{di}(x) 
  \gamma_0 \psi_{di}(x) \nonumber \\
  H_{I5} & = &  {i \over 2} \sum_i \int \!  d^3 x \, [d_u \bar{\psi}_{ui}(x) 
  \sigma_{\alpha \beta} \gamma_5 \psi_{ui}(x) 
        +  d_d \bar{\psi}_{di}(x) 
  \sigma_{\alpha \beta} \gamma_5 \psi_{di}(x)] F^{\alpha \beta}(x).
\end{eqnarray}
In the above $A^{\mu}$ and $F^{\mu \nu}$ are now understood to be quantized fields.
The evaluation of the diagrams we consider, shown in Fig.~\ref{1ab}, 
requires a treatment of the quark propagator, which when spherical symmetry
is present can be written as a partial-wave expansion in terms of $\kappa$.
This can be done with simple modifications of a numerical approach developed 
for atomic physics described in Ref.~\cite{fbs}.  In that paper a finite basis 
set using piecewise polynomials was applied to solving the Dirac equation for 
atoms. To discretize the spectrum the atoms were taken to be in a well with 
radius much larger than the size of the atom, with MIT bag boundary conditions 
applied at that radius. A typical application for atoms generates 50 positive-energy and 50 negative-energy states for each value of $\kappa$, with the 
advantage that the first few positive energy-states accurately reproduce the 
known positive-energy bound states. All that is needed to adapt the code to 
the present problem is to change units and to eliminate the potential, as 
there is now no nucleus, and the screening from the other quarks is treated 
perturbatively. We emphasize that in this approach the actual excited states 
of the neutron play no role: we are simply evaluating Feynman diagrams in a 
simple bound state treatment of the neutron, using an accurate numerical method to 
evaluate quark propagators. This point will be returned to below when we note 
that in some terms the $\kappa=1$ part of the quark propagator vanishes when 
negative- and positive-energy states cancel each other, an effect which would be missed 
if we worked with only positive-energy states. 

\begin{figure}[b]
  \begin{center}
    \includegraphics[width=7cm]{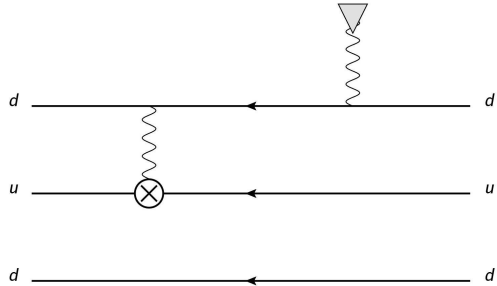}
  \end{center}
  \caption{Diagram \protect\ref{1ab}(a) for a particular assignment of quark flavors.}
  \label{d5}
\end{figure}

Fig.~\ref{d5} shows a part of  Fig.~\ref{1a} that involves $d_u$ and two
factors of $q_d$, with an energy shift we call $\Delta E_{a1}$.  We discuss its evaluation in detail to illustrate how both 
the neutron and the atomic calculations are carried out.
The photon and quark propagators are treated as follows. The photon part of the diagrams
arising from a contraction of $H_{I1}$ and $H_{I5}$ is 
\begin{equation}
  \langle 0 | T( A_{\mu}(x) F_{\alpha \beta}(y)) | 0 \rangle,
\end{equation}
which can be obtained from the usual Feynman gauge propagator
\begin{equation}
  \langle 0 | T( A_{\mu}(x) A_{\beta}(y)) | 0 \rangle = -i g_{\mu \beta} 
  \int {d^4 k \over (2\pi)^4} {e^{-i k \cdot (x-y)} \over k^2 + i \delta}
\end{equation}
by differentiation. Note that the time component of the derivative operator 
acts on the theta functions in the time-ordering operator, but because the 
equal-time commutator of photon fields vanishes one has 
\begin{equation}
  \langle 0 |\, T( A_{\mu}(x) F_{\alpha \beta}(y)) | 0 \rangle = 
  \int {d^4 k \over (2\pi)^4} (g_{\mu \beta} k_{\alpha} - g_{\mu \alpha} k_\beta)
  {e^{-i k \cdot (x-y)} \over k^2 + i \delta}.
\end{equation}

The quark propagator in any time-independent external field, here the static
well, can be expressed as
\begin{eqnarray}
  \langle 0 | T(\psi(x) \bar{\psi}(y))|0 \rangle & = & i S_F(x,y) \nonumber \\
  & = & i \int {dE \over 2\pi} e^{-iE(x_0-y_0)} S_F(\vec x, \vec y; E) \nonumber \\
  & = & i \int {dE \over 2\pi} e^{-iE(x_0-y_0)} \sum_m { \psi_m(\vec x) 
  \bar{\psi}_m(\vec y) \over E - \epsilon_m \pm i \delta}
\end{eqnarray}
with a factor $i \delta$ for positive-energy states and $-i \delta$ for 
negative-energy states. The infinite sum over both positive- and negative-energy states is approximated as a finite sum using the finite basis set 
techniques described above.

Using these forms for the propagators Fig. 2 corresponds to the energy shift
\begin{eqnarray}
 \Delta E_{a1} = -2 q_d^2 d_u^{} E_{\rm{ext}} \int d^3x d^3y d^3z 
 \int\frac{d^3k}{(2\pi)^3} \sum_m\frac{1}{\epsilon_g - \epsilon_m} 
 \frac{e^{i \vec k \cdot (\vec x - \vec z)}}{\vec{k}^2} k_{\alpha} \nonumber \\
 \langle n |:\psi_d^{\dagger}(\vec x) \gamma_0 \gamma_{\beta} \psi_m(\vec x)
 \psi_m^{\dagger}(\vec y) y_3 \psi_d(\vec y) \psi_u^{\dagger}(\vec z) \gamma_0
 \sigma^{\alpha \beta} \gamma_5 \psi_u(\vec z):|n \rangle.
\end{eqnarray}
At this point we make the approximation of setting $\beta=0$, 
which corresponds to accounting for internal Coulomb fields only 
and ignoring internal magnetic fields. We note that in the atomic case, Ref. 
\cite{Lindroth} did account for magnetic terms and showed they were extremely 
small. For the neutron we will find very small results for the Coulomb term, 
and while in this more relativistic case the magnetic terms may not be as 
suppressed as they are in a heavy atom, they should also be very small. The integral over $d^3k$ then
can be written as a gradient of the standard integral that gives 
$1/|\vec x - \vec z|$. This was dealt with in two ways. In the first,  
a partial integration was carried out, which gives
\begin{eqnarray}
 \Delta E_{a1} = \frac{2 q_d^2 d_u^{} E_{\rm{ext}}}{4\pi} 
 \int {d^3x d^3y d^3 z \over |\vec x - \vec z|} 
 \sum_m {1 \over \epsilon_g - \epsilon_m}  \nonumber \\
 \langle n |:\psi_d^{\dagger}(\vec x) \psi_m(\vec x)
 \psi_m^{\dagger}(\vec y) y_3 \psi_d(\vec y) \vec \nabla_z \cdot 
 ( \psi_u^{\dagger}(\vec z) \gamma_0
 \vec \Sigma \psi_u(\vec z)):|n \rangle.
\end{eqnarray}
A useful identify for evaluating the divergence term is
\begin{equation}
 \vec \nabla \cdot \bar{\psi}_m(\vec x) \vec \Sigma \psi_n(\vec x) 
  = \frac{1}{x^2}\sum_{JM} I_{JM}(-\kappa_n m_n,\kappa_m m_m) 
    Y_{JM}(\hat{x}) S_{mn}(x),
\end{equation}
with
\begin{equation}
  S_{mn}(x) = -(g_m(x) g_n(x))' + (f_m(x) f_n(x))' 
  -{\kappa_m + \kappa_n \over x} (g_m(x) g_n(x) + f_m(x) f_n(x))
\end{equation}
and
\begin{equation}
I_{JM}(\kappa_i \mu_i, \kappa_j \mu_j) = \int d \Omega Y_{JM}(\Omega) 
\chi^{\dagger}(\kappa_i \mu_i) \chi(\kappa_j \mu_j).
\end{equation}
The second evaluation did not use the partial integration, but applied the gradient operator directly to the $1/|\vec x - \vec z|$ term. These two approaches 
gave the same numerical final result, but it should be noted that in general care is needed with partial integration in the static well model.

In evaluating the diagram, one encounters a sum over all possible magnetic quantum numbers of quarks in the initial and final neutron states, along with those of the 
intermediate quark propagator (which in fact can only have $\kappa=1,-2$ from selection rules). In the atomic case, these magnetic-substate sums are always complete since we sum over the filled shells of the core, and standard identities of Racah algebra involving products of 3$j$ symbols can be used to reduce the angular portion of the calculation. Because three quarks in a neutron do not form filled shells, the sums in this case are not complete and such identities cannot be immediately applied. We therefore adopted two approaches. The first simply evaluates and sums the products of the three 3$j$ symbols that enter for the various combinations of magnetic quantum numbers---a `brute-force' approach. In the second, which will be described in detail elsewhere, the neutron wave function (\ref{neutronwf}) was rewritten in terms of Clebsch-Gordan coefficients and a fractional parentage decomposition for the spin couplings of the quarks, after which the standard identities of Racah algebra could be applied after all.  Once again, we obtained complete agreement between the two different approaches.

Now defining 
\begin{equation}
  R_{ij}(x) = g_i(x) g_j(x) + f_i(x) f_j(x),
\end{equation}
and
\begin{equation}
  r_{ij} = \int_0^R \! dx \, x R_{ij}(x),
\end{equation}
the diagram of Fig. 2 gives
\begin{eqnarray}
 \Delta E_{a1}
 & = & \frac{2}{243}\alpha d_u  
     E_{ext}\sum_{n_m}^{\kappa_m=1}
     \frac{r_{mg}}{\epsilon_g - \epsilon_m}
     \int_0^R dx \int_0^R dz
     \frac{r_<}{r_>^2} R_{gm}(x) S_{gg}(z) \nonumber \\
     & + & \frac{8}{729}\alpha d_u  
     E_{ext}\sum_{n_m}^{\kappa_m=-2}
     \frac{r_{mg}}{\epsilon_g - \epsilon_m}
     \int_0^R dx \int_0^R dz
     \frac{r_<}{r_>^2} R_{gm}(x) S_{gg}(z),
\label{dEa2}
\end{eqnarray}
which leads to 
\begin{equation}
d_n^{(2)}(a1) = 1.50 \times 10^{-4} d_u.
\end{equation}
An interesting feature of  the $\kappa=1$ part of $\Delta E_{1a}$ is that
it vanishes, so the result above comes entirely from the $\kappa=-2$ term. 
If only positive energy states are considered, a result on the order of
the $\kappa=-2$ term is found, but when negative energy states are
included an exact cancellation takes place. The vanishing of the $\kappa=1$ term can be
demonstrated analytically \cite{Mohr}. Were we to instead treat
the calculation by saturating the sum with actual excited neutron p-states we would
again obtain a nonzero result. The negative energy states doing the cancellation
of course involve the quark sea, here treated perturbatively. 

Similar manipulations apply to the other combinations of quark lines in Fig.~\ref{1ab},
and the final formulas are, keeping only those $\kappa=1$ terms that are nonzero,
\begin{equation}
 \Delta E_a
 =  -\frac{8}{729}\alpha \left( 10d_d - d_u \right) 
     E_{ext}\sum_{n_m}^{\kappa_m=-2}
     \frac{r_{mg}}{\epsilon_g - \epsilon_m}
     \int_0^R dx \int_0^R dz
     \frac{r_<}{r_>^2} R_{gm}(x) S_{gg}(z)
  \end{equation}
  and   
 \begin{equation}
 \Delta E_b 
 =  -\frac{8}{81}\alpha \left( d_u - d_d \right)
     E_{ext}\sum_{n_m}^{\kappa_m=1}
     \frac{r_{gm}}{\epsilon_g - \epsilon_m} 
     \int_0^R dx \int_0^R dz
     \frac{1}{r_>} R_{gg}(x) S_{mg}(z),
\label{dEb}
\end{equation}
which when evaluated numerically gives our result for the effect of the internal electric fields
in the neutron,
\begin{equation}
 d_n^{(2)} =  2.51 \times10^{-4}d_u - 1.60 \times10^{-3}d_d
 \,.
\label{dn2}
\end{equation}

While it is of note that the contribution from the down quarks dominates that 
from the up quarks by an order of magnitude, the overall contribution from 
internal-field effects is significantly suppressed relative to the first-order 
neutron edm $d_n^{(1)}$ given in Eq.~\eqref{dn1}. No effect corresponding to 
the dramatic violation of Schiff's theorem in heavy atoms is found, a result 
which we discuss further in the conclusions.

\section{Atomic Calculations}

Turning now to the atomic case, we first note that manipulations with 
second-order perturbation theory are generally made that involve replacing the 
electric field felt by an electron in an atom, which is comprised of a 
one-body term coming from the nucleus and a sum of two-body terms coming from 
the other electrons, with a commutator. After this is done, a cancellation 
with first-order perturbation theory takes place, and one is left with an 
effective CP-violating one-body operator, the `Sandars operator' \cite{Sandars}, given by the commutator
\begin{equation}
 [\gamma_0 \vec \Sigma \cdot \vec p, \vec \alpha \cdot \vec p] 
 = 2 \gamma_0 \gamma_5 \vec p\,^2 \,.
\label{sandarsop}
\end{equation}
This commutator would vanish if the factor $\gamma_0$ in the first term were not present. 
This has the advantage both of showing that Schiff's theorem holds in nonrelativistic 
systems, when $\gamma_0$ is effectively close to $1$, and also of allowing 
one to work with only a one-body operator. We note that interesting discussions of
alternative ways of deriving a CP-violating one-body operator have been given in Ref. \cite{Lindroth},
where calculations for paramagnetic atoms are presented, and more recently in Ref. \cite{Haxton},
where the theory of the edm of diamagnetic atoms is treated in detail.

If the original Sandars form is used in a first order MBPT calculation, a useful feature
of the CH potential is that the result turns out to be 
exactly equal to the sum of the dominant three terms of the field theory calculation to 
be presented below. This is a special property of the CH potential, and use of other
local potentials, while still dominated by the same three terms, gives results that
differ from first order MBPT.  Two of the three terms have already been encountered
in the neutron calculation, but for the atomic problem an 
electric field arising from the nucleus is now present, so that we need to 
include an 
additional interaction Hamiltonian,
\begin{equation}
  H_{I4}  =  -i d_e q_e \int d^3  x \bar{\psi}(x) \sigma_{0i} \gamma_5 
  \psi(x) {x_i \over 4 \pi r^2}\left[{Z(r) \over r} - Z'(r)\right] \,.
\label{nuclearterm}
\end{equation}
The interaction Hamiltonians used in the neutron calculation all have atomic 
analogs, which can be taken from the previous section by eliminating the sums 
over color, dropping terms involving the down quark, and replacing $q_u$ and 
$d_u$ with $q_e$ and $d_e$, respectively. We retain the convention of keeping 
the valence electron spin up (the core of course has zero spin), but instead 
of presenting the coefficient of $-E_{\rm{ext}}$, we give results in terms of 
$R$ as defined in Eq.~\eqref{Rdef}; that is, we also pull out a factor $d_e$. We 
illustrate the calculation with the case of cesium with the CH potential, but 
give results for thallium and the full set of alkalis in Table~\ref{tab:Ratom}. 
The MBPT result for cesium with the standard form of the Sandars operator given above 
is an enhancement factor of 154.657. We note that this is only first-order 
perturbation theory, and the difference with more complete MBPT calculations 
mentioned in the introduction is to be expected. 

The same Feynman diagrams are present as for the neutron, but in the atomic 
case there is a sum over core states as the valence electron interacts with the 
electrons in the core. Diagrams analogous to Fig.~\ref{1ab} in which the photon interacts between two core states do not contribute to the atomic edm, because the core has total spin zero.  In general, one external line in any Feynman diagram must correspond to a valence electron for angular symmetry reasons.

\begin{figure}[t]
  \begin{center}
    \includegraphics[width=7cm]{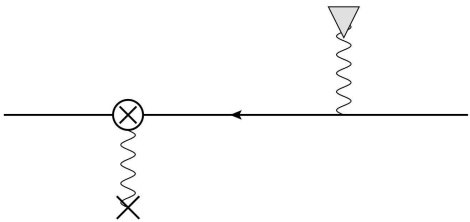}
  \end{center}
  \caption{A diagram representing the interaction of the valence electron's 
    edm with the Coulomb field of the nucleus, represented by an $X$.}
  \label{nucl}
\end{figure}

The presence of the additional Hamiltonian associated with the nuclear Coulomb
field (Eq. \ref{nuclearterm}) requires evaluation of the diagram of Fig.~\ref{nucl}. The diagrams of Fig.~\ref{1ab} are 
now to be thought of as having the upper line be the valence electron and the 
middle line one of the core states; a sum over all possible core states is 
implied for that line.  There is also a second set of diagrams in which the top line is a core line and the middle line is the valence state. For both sets, the roles of external core and valence states on the left-hand side of the diagram only can also be interchanged (yielding a diagram in an `exchange' rather than a `direct' configuration).  The lower ``spectator quark'' line is eliminated. A complication is the fact that there can be terms with nonvanishing energy differences flowing through the photon propagator (the neutron does involve photon exchange between quarks, but all energies are identical in that case).

The calculations are 
quite similar to MBPT, with three exceptions. 
First, as mentioned above, most atomic calculations carry out a rearrangement of the Hamiltonian 
that builds in Schiff's theorem as much as possible, leaving only a one-body 
operator \eqref{sandarsop} that vanishes in the nonrelativistic limit, and this CP-violating operator is then treated in MBPT.
Here, while the 
interaction with the nucleus is also a one-body operator, many of our diagrams are of two-body type, so the enhancement factor arises in a different way. 
Second, the electron propagators contain both positive- and negative-energy 
states. 
Finally, while the sums over positive-energy states in MBPT are generally
arranged in such a way as not to include both occupied core states and excited states in a single sum, in our field-theoretic approach the
electron propagator includes core states along with excited states (and negative-energy states). However, as long as a complete set 
of diagrams is included, cancellations will occur between different diagrams that
effectively enforce this restriction where appropriate. This property was checked as part of the verification of our calculation.

Because we are not building in Schiff's theorem, an important contribution
arises from lowest order, which parallels the lowest-order neutron result.
Evaluating the effect of  $H_{I3}$ for an atom with a valence $s_{1/2}$ state gives the enhancement factor
\begin{equation}
  R_0 =  \int_0^{\infty} \! dr \, [g_v^2(r) + \frac13 f_v^2(r)] \,,
\label{zerothorder}
\end{equation}
which, as mentioned before, is very close to $1$. To exhibit Schiff's
theorem for a light alkali atom then clearly requires finding terms  from the nuclear and one-photon-exchange diagrams that sum to a result close to $-1$.

An interesting feature of the field-theoretic approach is that one 
automatically generates diagrams involving radiative corrections on either the 
edm or external field vertex, with a representative diagram shown in Fig.~\ref{self}.
\begin{figure}[b]
  \begin{center}
    \subfigure[]{\label{self}\includegraphics[width=7cm]{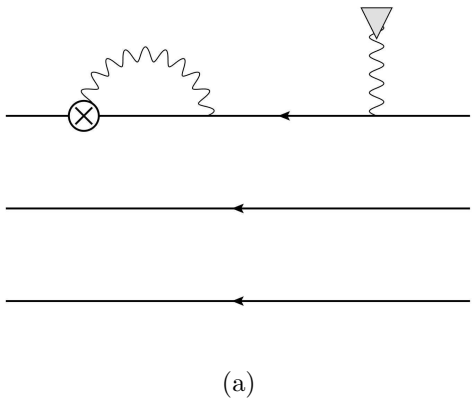}}
  \end{center}
  \caption{Self-energy type correction to edm}
  \label{self}
\end{figure}
 The presence of such terms was noted in 
Ref. \cite{Lindroth}. Techniques to evaluate QED radiative corrections in the 
presence of perturbing potentials have been developed in Ref. \cite{radpert}, 
and these corrections were found to have at least a factor of $\alpha$ 
suppression. For that reason, we do not treat them here. 

Figure~\ref{nucl} leads to
\begin{equation}
  R_{\rm nuc} = {2 \over 3} \alpha \sum_m^{\kappa_m =-\kappa_v} {d_{vm} r_{mv} 
  \over \epsilon_v - \epsilon_m}\,,
\label{rnuc}
\end{equation}
with
\begin{equation}
  d_{ij}  = \int_0^{\infty} \! dr \, [g_i(r) g_j(r) - f_i(r) f_j(r)]\left[{Z(r) \over r^2}
  - {Z'(r) \over r}\right] \, .
\end{equation}
Since no account of the electric field coming from other electrons has yet been made, a very different result
from the Sandars form of $154.657$ is found,
\begin{equation}
  R_{\rm nuc} = -80.934.
\end{equation}
The bulk of this comes from the $6p_{1/2}$ state, $-86.354$, with the next two 
most important contributions being from the $5p_{1/2}$ state,
7.128, and the $7p_{1/2}$ state, $-1.182$. 

Next we turn to the photon-exchange diagrams.  The first diagram arises from Fig.~\ref{1b} when the upper external lines are valence states and the middle external lines are core states.  In this case, the external field interacts in the usual way with the valence electron, which subsequently has its edm interact with a core electron through photon exchange. It is this term that dominates and acts to cancel the nuclear term, as will be shown below.  Other, smaller terms arise from Fig.~\ref{1b} when the external core and valence lines are interchanged in various ways, and from Fig.~\ref{1a}.  All these cases intrinsically involve two-body interactions, which are not particularly complicated to deal with. As with the neutron, we make the approximation of keeping only the $\mu=0$ component of the photon propagator. The full expression for Fig.~\ref{1b}, including all combinations of external core and valence lines, is
\begin{eqnarray}
  R_a & = & -2 \alpha \sum_{ma} {z_{vm} \over \epsilon_v - \epsilon_m} 
  \int {d^3 x d^3 z \over |\vec x - \vec z|} \psi^{\dagger}_a(\vec x)
  \psi_a(\vec x) \vec \nabla_z \cdot (\bar{\psi}_m(\vec z) \vec \Sigma 
  \psi_v(\vec z))  \nonumber \\
  & - & 2 \alpha \sum_{ma} {z_{am} \over \epsilon_a - \epsilon_m} 
  \int {d^3 x d^3 z \over |\vec x - \vec z|} \psi^{\dagger}_v(\vec x)
  \psi_v(\vec x) \vec \nabla_z \cdot (\bar{\psi}_m(\vec z) \vec \Sigma 
  \psi_a(\vec z)) \nonumber \\
  & + & 2 \alpha \sum_{ma} {z_{am} \over \epsilon_a - \epsilon_m} 
  \int {d^3 x d^3 z \over |\vec x - \vec z|}e^{i(\epsilon_v - \epsilon_a)
  |\vec x - \vec z|} \psi^{\dagger}_v(\vec x)
  \psi_a(\vec x) \vec \nabla_z \cdot (\bar{\psi}_m(\vec z) \vec \Sigma 
  \psi_v(\vec z)) \nonumber \\  
  & + & 2 \alpha \sum_{ma} {z_{vm} \over \epsilon_v - \epsilon_m} 
  \int {d^3 x d^3 z \over |\vec x - \vec z|}e^{i(\epsilon_v - \epsilon_a)
  |\vec x - \vec z|} \psi^{\dagger}_a(\vec x)
  \psi_v(\vec x) \vec \nabla_z \cdot (\bar{\psi}_m(\vec z) \vec \Sigma 
  \psi_a(\vec z)) \,
\label{r125B}
\end{eqnarray} 
where $v$ is the valence state, the states $a$ are core states, and the sum over $m$ is over a complete set of states (positive- and negative-energy) arising from the intermediate electron propagator. To reach this form a partial integration has been carried out, but, as with the neutron, we also did the calculation without doing this and obtained the same result to high numerical accuracy. We have dropped two terms that do not 
depend on $v$ because they vanish by angular symmetry, as mentioned earlier, which is also the case for the second term in the above expression. The dominant term is the 
first, which can be identified with the charge distribution of the core interacting with the edm of the valence electron; the other two terms play a much smaller numerical role. 

After angular reduction, the dominant first term of $R_a$ has the form
\begin{equation}
  R_{a1} = {2 \alpha \over 3} \sum_{[m][a]}^{\kappa_m = -\kappa_v} 
  {r_{vm} \over \epsilon_v - \epsilon_m}(2j_a+1) \int_0^{\infty} dx
  \int_0^{\infty}dz {1 \over {r_>}} R_{aa}(x) S_{mv}(z).
\end{equation}
For cesium this has the value
\begin{equation}
  R_{a1} = 234.590,
\end{equation}
again dominated by the $6p_{1/2}$ intermediate state, which contributes 222.327.
The accumulated sum so far is now 154.657, exactly equal to the 
Sandars form result. We note that this pattern is quite different from
Ref.~\cite{Lindroth}, where the nuclear term dominates. However, the 
meaning of the nuclear term is different in that work,
involving a factor $\gamma_0 -1$ not present in our approach, so the 
disagreement is to be expected. The other terms in Eq.~\eqref{r125B}, as well as those arising from Fig.~\ref{1a}, are all much smaller (or zero), contributing to $R$ at 
the level of 0.2 or less for cesium, and we will not present their values.

While of course not of interest for a practical determination of $d_e$, the
case of lithium shows how this approach implements Schiff's theorem. The
cesium results of (1, -80.934, 234.590) go to (1, -8.266, 7.270), summing to 0.004.
Three features of the lithium calculation are of note. The first is that the 
small terms we have ignored for heavy atoms, while still summing to a small value, 
can be as large as 0.7 in magnitude individually. The second is that we found that the 
approximation of dropping the exponential terms, the no-retardation 
approximation, was in some cases a poor one for a particular diagram, though 
this had only a small effect on the total answer. The third is that using potentials other than CH fails to exhibit the almost exact cancellation shown above for lithium. In 
such a case presumably higher-order terms would improve the cancellation, but 
of course atoms that do not have a large value of $R$ are not of interest for 
putting limits on $d_e$. In Table 1 we present the three dominant contributions 
to all alkali atoms and thallium along with their sums. We again emphasize 
that these $R$ values correspond to first-order MBPT, and are included here to show the
Schiff suppression for light atoms and the Sandars enhancement for heavy atoms, and not
to represent high accuracy results.

\begin{table}
\caption{\label{tab:Ratom}Atomic edm enhancement factors given by the present formalism using a CH potential (see text).  Notation: $R_0$ is the lowest-order term, Eq.~\protect\eqref{zerothorder} for a valence state, $R_{\rm nuc}$ is the nuclear term, Eq.~\protect\eqref{rnuc}, and $R_a$ is the dominant photon-exchange term arising from the first term in Eq.~\protect\eqref{r125B}.}
\begin{ruledtabular}
\begin{tabular}{cdddd}
Atom &  \multicolumn{1}{c}{$R_0$} & \multicolumn{1}{c}{$R_{\rm nuc}$} & \multicolumn{1}{c}{$R_a$} & \multicolumn{1}{c}{Sum} \\
\hline
Li ($2s_{1/2}$) & 1.000 & -8.266  & 7.270  & 0.004 \\
Na ($3s_{1/2}$) & 1.000 & -31.875 & 31.314 & 0.439  \\
K  ($4s_{1/2}$) & 1.000 & -66.220 & 68.807 & 3.588  \\
Rb ($5s_{1/2}$) & 1.000 & -110.428& 143.160 & 33.732 \\
Cs ($6s_{1/2}$) & 1.000 & -80.934 & 234.590 & 154.657 \\
Fr ($7s_{1/2}$) & 1.000 & 701.730 & 364.161 & 1066.891 \\
Tl ($6p_{1/2}$) & -0.333 & -573.199 & -219.132 & -792.665 \\
\end{tabular}
\end{ruledtabular}
\end{table}

\section{conclusions}

The field-theory approach given here for calculating the edm of the neutron and 
paramagnetic atoms, arising from their constituents having edms, has 
shown firstly that for the neutron, outside of a less than 20 percent 
normalization shift from the nonrelativistic result, that electromagnetic 
corrections are small. We attribute this smallness to the fact that unlike 
atoms, where the energy shift associated with internal electric fields is of the order 
of the energy scale of atomic transitions, in the neutron the electromagnetic 
energy shifts are of order 1 MeV, while transition energies in the static well model are hundreds of MeV.  For the neutron, the diagrams in Fig.~\ref{1ab} are suppressed by the ratio of the internal electromagnetic interaction energy to these excitation energies [see Eqs.~\eqref{dEb}--\eqref{dn2}], giving a result three orders of magnitude smaller than the lowest-order result \eqref{dn1}.  Thus, although the quarks are highly relativistic within the model, the effect of the internal electromagnetic interactions on the edm is quite different from that found in heavy atoms, where the same class of diagrams yields large enhancement factors.

Secondly, the field-theoretic approach has provided a different way of calculating the enhancement 
factor $R$, not involving the Sandars \cite{Sandars} operator \eqref{sandarsop}, although both one and two-body effects have to be evaluated.

\begin{figure}[b]
  \begin{center}
    \subfigure[]{\label{glu1}\includegraphics[width=7cm]{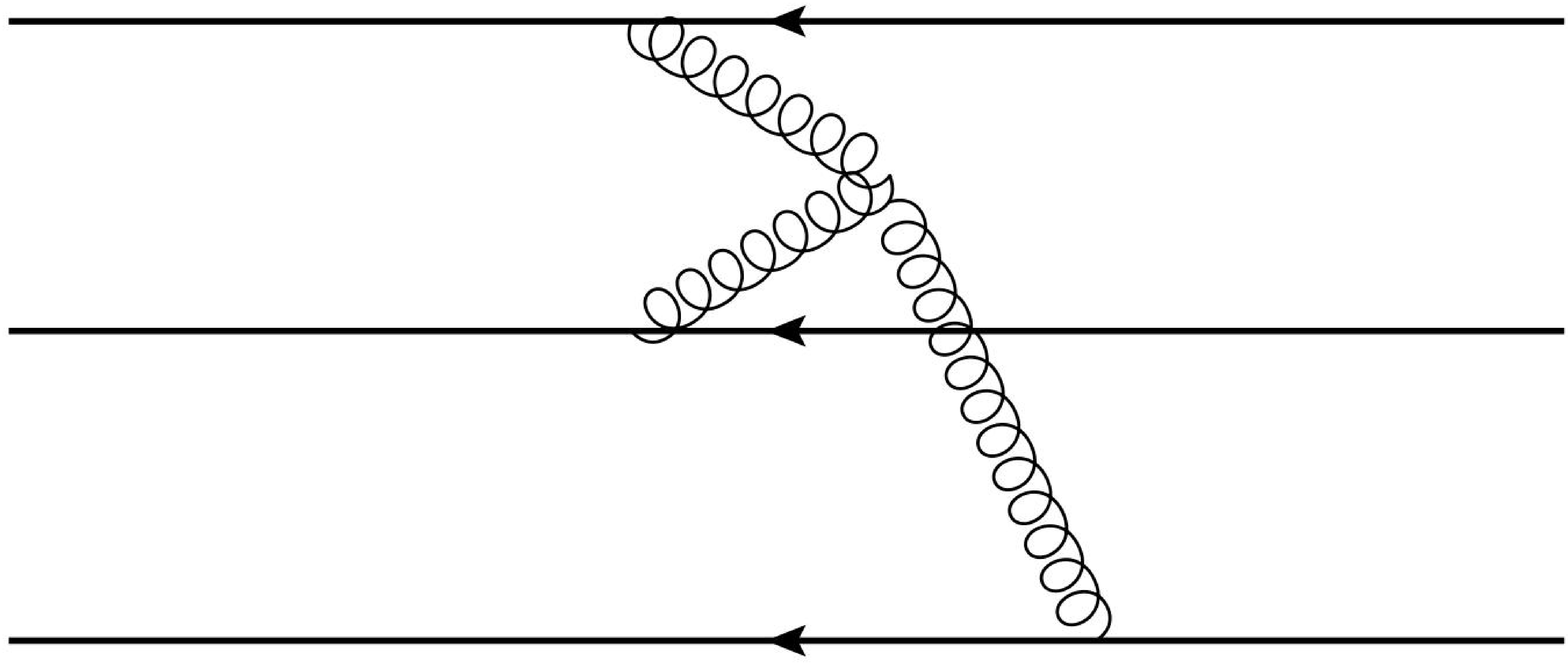}}
    \quad
    \subfigure[]{\label{glu2}\includegraphics[width=7cm]{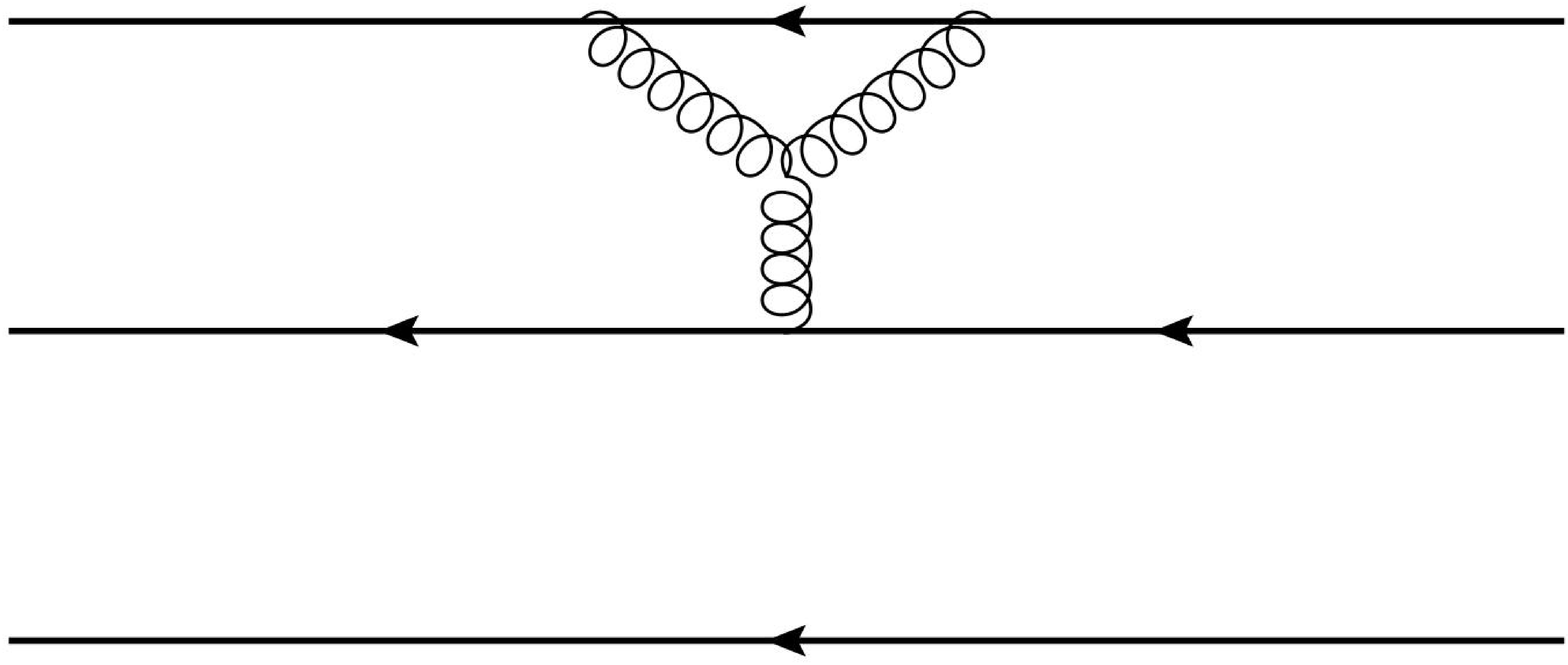}}
  \end{center}
  \caption{Representative Weinberg 3-gluon diagrams contributing to $d_n$.}
  \label{3glu}
\end{figure}

We now discuss possible extensions of this approach for the case of the 
neutron. We note that most treatments of the neutron edm involve naive
dimensional analysis, sum-rule techniques or chiral perturbation theory, as 
described in Ref.~\cite{Roberts}. The approach used here, while based on a very 
simple static-well model, could in principle be used to provide an alternative method of 
calculation. A particularly interesting application could be an evaluation 
of the effect of Weinberg's CP-violating 3-gluon vertex \cite{Weinberg} on $d_n$. This has
been treated by Bigi and Uraltsev \cite{Bigi} and  
Demir, Pospelov and Ritz \cite{Pospelov}.   
In the present approach one would  evaluate the two diagrams shown in 
Fig.~\ref{3glu}, with the understanding that the 3-gluon vertex is the source of CP-violation and that an interaction with the external electric field is to be added in all lines.

For atoms, we have demonstrated a different way of showing Schiff's theorem 
for nonrelativistic atoms and calculating the enhancement factor for 
relativistic atoms. However, the field theory here corresponds to only first-order MBPT, and it would be difficult to extend it even to second order, as 
ladder and crossed-ladder diagrams would have to be evaluated. The leading effect of these diagrams would be expected
to reproduce second order MBPT, and arguments were given in Ref.  \cite{Lindroth} that subleading terms involving negative energy states
are numerically unimportant. Probably the 
most interesting extension of this work in the atomic case would be to 
radiative corrections, but because of their smallness there seems no need at 
present for the considerable effort such calculations entail.

\begin{acknowledgments}
The work of J.G. and J.S. was supported in part by NSF grant PHY-1068065. We thank Peter Mohr for helpful conversations.
\end{acknowledgments}


\begin{references}
\bibitem{Roberts} {\it Lepton Dipole Moments}, ed. B. Lee Roberts and William J. Marciano, (World Scientific, Singapore, 2010).
\bibitem{Schiff} L. Schiff, Phys. Rev. {\bf 132}, 2194 (1963).
\bibitem{Sandars} P.G.H. Sandars, Phys. Lett. {\bf 14}, 194 (1965); Phys. Lett. {\bf 22}, 290 (1966).
\bibitem{cs} H.S. Nataraj, B.K. Sahoo, B.P. Das, and D. Memherjee, Phys. Rev. Lett. {\bf 101}, 033002 (2008);
V.A. Dzuba and V.V. Flambaum, Phys. Rev. A{\bf 80}, 062509 (2009).
\bibitem{tl} Z.W. Liu and H.P. Kelly, Phys. Rev A {\bf 45}, R4210 (1992). 
\bibitem{Lindroth} A somewhat different implementation of
field theory to the atomic case was previously given by E. Lindroth, B.W. Lynn, and P.G.H. Sandars, J. Phys. B {\bf 22}, 559 (1989).
\bibitem{MIT} A. Chodos, R.L. Jaffe, K. Johnson, C.B. Thorn, and V.F. Weisskopf, Phys. Rev. D {\bf 9}, 3471 (1974).
\bibitem{Lee} See chapter 20 of {\it Particle Physics and Introduction to Field Theory}, T.D. Lee, (Harwood Academic Publishers, London, 1988).
\bibitem{Furry} W. Furry, Phys. Rev. {\bf 81}, 115 (1951).
\bibitem{Sucher} J. Sucher, Phys. Rev. {\bf 107}, 1448 (1957).
\bibitem{Rev} J. Sapirstein, Rev. Mod. Phys. {\bf 70}, 55 (1998).
\bibitem{Commins} E.D. Commins, J.D. Jackson, and D.P. DeMille, Am. J. Phys. {\bf 75}, 532 (2007).
\bibitem{fbs} J. Sapirstein and W.R. Johnson, Phys. Lett. B{29}, 5213 (1996).\bibitem{Mohr} P.J. Mohr, private communication.
\bibitem{Haxton} C.-P. Liu, M.J. Ramsey-Musolf, W.C. Haxton, R.G.E. Timmermans, and E.L. Dieperink, Phys. Rev. C {\bf 76}, 035503 (2007).
\bibitem{radpert} S.A. Blundell, K.T. Cheng, and J. Sapirstein, Phys. Rev. A {\bf 55}, 1857 (1997).
\bibitem{Weinberg} S. Weinberg, Phys. Rev. Lett. {\bf 63}, 2333 (1989).
\bibitem{Bigi} I.I. Bigi and N.G. Uraltsev, Nucl. Phys. B{\bf 353}, 321 (1991).
\bibitem{Pospelov} D.A. Demir, M. Pospelov and A. Ritz, Phys. Rev. D{\bf 67}, 015007 (2003).
\end{references}
\end{document}